# Steady state rheological behaviour of multi-component magnetic suspensions

**Laura Rodríguez-Arco,**[a] **Modesto T. López-López,\***[a] **Pavel Kuzhir**[b] **and Juan D.G. Durán**[a]

[a] *Departamento de Física Aplicada, Facultad de Ciencias, Universidad de Granada, Avda. Fuentenueva s/n, 18071 Granada, Spain. Fax: +34 958 243214; Tel: +34 958 240076;E-mail:modesto@ugr.es*

[b] *CNRS, U.M.R. 7336, Laboratoire de Physique de la Matière Condensée, Université de Nice,*

*Parc Valrose, 06108 Nice, France. Fax: +33 492076536; Tel: +33 492076313;E-mail:kuzhir@unice.fr*

**Abstract**

In this paper we study the rheological behaviour (in the absence of magnetic field and upon its application) of multi-component magnetic suspensions that consist of a mixture of magnetic (iron) and non-magnetic (PMMA) particles dispersed in a liquid carrier. These suspensions exhibit considerably higher viscosity and yield stress in the absence of magnetic field than single-component suspensions of the same solid fraction, as a consequence of the adsorption of the iron particles on the PMMA ones. The adsorbed layer of iron particles on the PMMA ones is observed through optical microscopy of dilute samples and confirmed by attenuated total reflectance. Microscopic observations also show that the resulting non-magnetic-core–magnetic-shell composites move upon magnetic field application and aggregate into particle structures aligned with the applied field. These structures, which consist of both types of particles, give rise to high values of the static and dynamic yield stresses upon field application. Actually, both quantities are much higher than those of a suspension with the same volume fraction of magnetic particles, and increase when the amount of non-magnetic ones increases. These trends are adequately predicted by a theoretical model that considers that the main contribution to the yield stress is the change of the suspension magnetic permeability when particle chains are deformed by the shear.

## 1. Introduction

Multi-component fluid systems have attracted much attention in the last decades, especially in the search for new tailor-made smart materials, whose properties can be controlled by changing the nature either of the different phases dispersed in the continuous matrix, or of this latter. The combination of different materials turns out to be very interesting because each of them provides with a specific feature to the whole blend. This is the case, for example, of multi-component mixtures of surfactants, whose interfacial properties are different from those of a solution involving a single one.[1] Furthermore, the mixture of different materials may lead to the enhancement of the properties of the whole blend with respect to those of the single components. An example of this synergic effect is the increase of both the yield stress and the current density for bidisperse electrorheological fluids that consist of large particles and a small fraction of fine particles dispersed in a liquid carrier.[2] Similarly, magnetic microparticles have been dispersed in a ferrofluid. In this case, the stability against aggregation and redispersion were better than those of a suspension constituted only by the magnetic microparticles.[3] Nevertheless, the inclusion of different phases is not only interesting because of the improvement of the technological-related properties. It is very interesting from the fundamental point of view too. As an example, it was recently reported the existence of magnetic field-induced repulsion (instead of attraction) between magnetic microparticles dispersed in a ferrofluid, even though their magnetic moments were in line.[4] This surprising phenomenon is due to the phase condensation of the ferrofluid carrier near the micron-sized particles and, as a result, the interactions between the microparticles could be controlled just by changing the concentration of the nanoparticles in the ferrofluid.[5]

The control of the interaction between the dispersed particles in suspensions is very important in fields such as self-assembly, which actually has taken advantage of multi-component systems too. For example, Lee et al.[6] have theoretically proved that the inclusion of bidisperse systems of spheres in block copolymers, allows controlling both the morphology of the polymer matrix and the spatial organization of the spheres. Something similar has been found by Cho et al. who prepared colloidal aggregates of nano- and micron-sized particles in water-in-oil emulsion droplets.[7] However, self-assembled structures can be obtained, not only by the use of different particle sizes, but also by the combination of particles of different nature. In this sense, it is particularly interesting the work recently published in *Nature*[8] in which it is reported the formation of symmetric magnetic superstructures due to the interactions between diamagnetic and paramagnetic particles dispersed in a ferrofluid.

The broad range of commercially available diamagnetic (non-magnetic) particles has resulted in other attempts to include them in the formulation of magnetic suspensions.[9] For example, inverse ferrofluids consist of micron-sized non-magnetic particles dispersed in a ferrofluid.[10-12] Similarly, López-López et al. prepared suspensions that consisted of iron microparticles stabilized by the addition of organoclay particles. They found that these suspensions developed, at low magnetic fields (up to $H \approx 10$ kA/m), higher yield stresses than suspensions of just magnetic (iron) microparticles (also known as magnetorheological (MR) fluids). The explanation given by the authors to this behaviour was a combined effect of the field-induced iron chains and a gel formed by the clay particles. The formation of such a gel gave rise to high yield stresses in the absence of magnetic field too.[13] Other authors [14-16] have pointed out in a similar direction by reporting that the addition of non-magnetic particles to conventional MR fluids led to an enhancement of their yield stress.

In the present paper we rigorously study the steady state rheological behaviour both in absence of magnetic field and upon its application, of a particular kind of multi-component fluid system. More specifically, the studied systems consist of two different particle populations which differ in both their magnetic behaviour and size. As a matter of fact the dispersed phase is composed of magnetic and non-magnetic microparticles with particle diameters of about 2.3 μm and 10 μm respectively. The non-magnetic particle volume fraction for the prepared suspensions is varied from 10 to 30 vol %, while the magnetic particle one is kept constant and equal to 10 vol %. The rheological behaviour of samples constituted just by magnetic or non-magnetic particles (i.e. single-component suspensions) is studied for comparison too. As will be shown, in the absence of field, the multi-component suspensions show higher values of the viscosity and the yield stress than the single-component ones. This makes evident that there must be some kind of interaction between both populations of particles and, for this reason, microscopic observations and attenuated total reflectance measurements were performed. The particle structures formed upon field application were also observed by microscopy in order to better understand the rheological behaviour in the presence of external magnetic fields. To end with, a theoretical model for the static yield stress is applied to these suspensions. In order to estimate the theoretical static yield stress, finite element method (FEM) simulations of the suspension magnetic permeability were also required.

## 2. Experimental

### 2.1. Materials

Iron and poly(methylmethacrylate) (PMMA) powders were purchased from BASF (HS quality) and Microbeads (Spheromers CA10) respectively. The first one consists of polydisperse spherical particles with a median particle size of $d_{50} = 2.3$ μm (particle diameter ranges from 0.5 to 3 μm) and an iron content of 97% (density = 7.5 g/cm$^3$). Its physicochemical properties are reported elsewhere.[17] PMMA powder is composed of particles with an average diameter of (9.9 ± 0.4) μm and purity of 99.5% (density = 1.2 g/cm$^3$). Mineral oil (Sigma Aldrich) was used as the liquid carrier for all the suspensions; its viscosity at 25 °C is (0.028 ± 0.001) Pa·s.

## 2.2. Suspension preparation

First of all, several suspensions of PMMA microparticles were prepared. For this purpose, the desired amounts of the solid powder were dispersed in the liquid carrier and the suspension was stirred both manually and mechanically in order to homogenize it. The volume fraction of PMMA ranged from 20 to 40 vol %. In a similar way, multi-component suspensions constituted by both iron and PMMA microparticles dispersed in mineral oil were prepared. Iron/PMMA volume fractions in mineral oil were as follows: 10/10, 10/20 and 10/30. These samples will be labelled from now on as S1(10:10), S2(10:20) and S3(10:30) respectively. Note that the resulting multi-component suspensions were as PMMA suspensions in which 10 vol % of the PMMA particles (with respect to the total volume) were substituted by the same amount of iron particles –i.e. the solid concentration of the multi-component suspensions ranged from 20 to 40 vol %, as it did in the above-mentioned PMMA suspensions. In addition, suspensions composed just of iron microparticles (10 and 14 vol %) were also prepared for comparison (samples S0(10:0) and S0b(14:0)).

## 3. Experimental results

### 3.1. Steady state rheological measurements in the absence of magnetic field

The rheological behaviour of the samples in the absence of magnetic field was firstly studied with the aim of gaining information on the zero-field state of the samples. The rheological characterization (at 25 °C) was performed with a Haake MARS III (Thermo Fisher Scientific, USA) rheometer. The measurement system geometry was a 3.5 cm diameter, parallel plate set. The gap between both plates was kept at 400 µm and the plate surface was rough in order to avoid wall slip. The reported quantities are those at the outer radial edge of the plate. All the samples were well redispersed by hand and by a vortex mixer prior to being placed in the measuring system. Before carrying out each measurement we subjected the samples to the following protocol: (i) pre-shear: 30 s of 150 s$^{-1}$ shear rate application; (ii) waiting time: sample at rest for 2 minutes. The reproducibility of the results was checked by performing several repetitions of the measurements; the obtained relative dispersion was always lower than 5 %.

In order to estimate the yield stress (i.e. the stress above which the suspension flows) we performed steady shear measurements. The so-called dynamic yield stress (related to the complete breakage of the existing structures in the suspension) can be obtained by a linear interpolation of the rheogram (i.e. shear stress, $\sigma$, plotted as a function of shear rate, $\dot{\gamma}$) to zero shear rate. This linear fit is usually performed at relatively high shear rates.[18] For this reason, all the samples were firstly subjected to a linear shear rate ramp (controlled rate, CR, measurements) from 20 to 300 s$^{-1}$. In these measurements, a particular value of the shear rate was imposed for at least 3 seconds and at the most for 30 seconds. Within this time range, if the quotient $(\Delta\sigma/\sigma)/\Delta t$ was smaller than 0.001 s$^{-1}$, the measurement at the given shear rate was stopped. In this formula, $\Delta\sigma$ is the difference between the highest and the lowest shear stresses at the applied shear rate, and $\Delta t$ the time during which the shear rate was maintained. In all cases, the average value of the shear stress was the one represented in the rheograms.

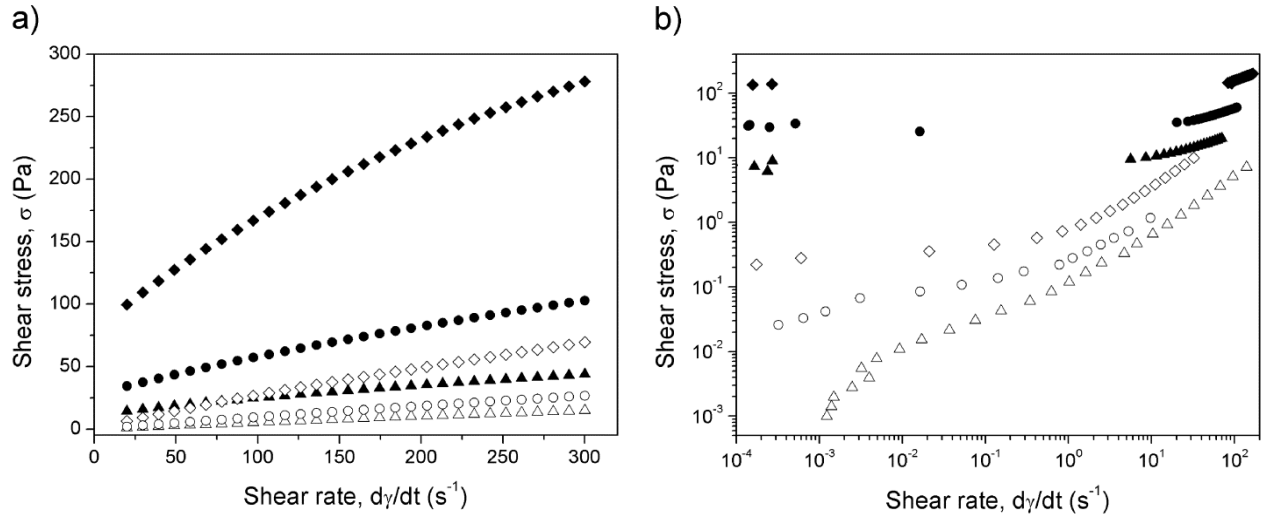

**Fig. 1** CR (a) and CS (b) rheograms for PMMA suspensions (open symbols) and for multi-component suspensions (full symbols). The different symbols correspond to different total volume fractions: △/▲: 20 vol %; ○/●: 30 vol %; ◇/◆: 40 vol %. Note that in the multi-component suspensions, the volume fraction of iron is 10 %.

Table 1 Characteristic rheological parameteres of the suspensions at zero-field.

| Particle volume fraction | Experimental plastic viscosity (Pa·s) | Krieger-Dougherty viscosity (Pa·s) | Deviation of viscosity from Krieger-Dougherty prediction (%) | Static yield stress (Pa) | Dynamic yield stress (Pa) |
|---|---|---|---|---|---|
| 20% PMMA | 0.05 ± 0.05 | 0.051 | 10 | 0.034 ± 0.009 | 1.17 ± 0.03 |
| 30% PMMA | 0.08 ± 0.05 | 0.078 | 5 | 0.124 ± 0.021 | 2.11 ± 0.08 |
| 40% PMMA | 0.20 ± 0.05 | 0.140 | 47 | 0.39 ± 0.07 | 8.12 ± 0.08 |
| 10% iron | 0.05 ± 0.05 | 0.037 | 23 | 5.59 ± 0.20 | 11.88 ± 0.24 |
| 14% iron | 0.06 ± 0.05 | 0.042 | 32 | 7.9 ± 0.9 | 16.09 ± 0.04 |
| 10% iron, 10% PMMA | 0.09 ± 0.05 | 0.051 | 70 | 9.0 ± 0.4 | 17.95 ± 0.13 |
| 10% iron, 20% PMMA | 0.21 ± 0.05 | 0.078 | 166 | 33.8 ± 1.0 | 40.8 ± 0.3 |
| 10% iron, 30% PMMA | 0.46 ± 0.05 | 0.140 | 228 | 137.9 ± 2.4 | 141.7 ± 1.6 |

The static yield stress (which corresponds to the breakage of the structures in their weakest point and therefore, to the onset of the flow) is usually quantified by extrapolating the value of the shear stress in the low shear rate pseudoplateau (around 0.1 s$^{-1}$) to zero-shear rate in double logarithmic scale rheograms.[18] For this reason, two kinds of measurements were performed to estimate it. First of all, the samples were subjected to a logarithmic stress ramp which allowed us to delimit the range of stresses in which the static yield stress could be found. Afterwards, we applied a linear stress ramp within the previously delimited range (controlled stress, CS, measurements). The measurement duration was set in a similar way to this previously described for CR measurements, but this time the quotient $(\Delta\dot{\gamma}/\dot{\gamma})/\Delta t$ was employed.

All these measurements were performed firstly for the PMMA suspensions and then for the multi-component suspensions. Figure 1 shows the rheograms obtained in the CR (a) and CS (b) modes for all these samples.

Concerning PMMA suspensions, as observed in Figure 1, their dynamic yield stress (intercept of the curves) is almost negligible and their viscosity (slope) is approximately independent of the shear rate. Note that the slope of the curves increases with the particle volume fraction. This reflects an increase of viscosity due to a higher particle content that makes hydrodynamic interactions more important.[18] However, a more detailed inspection reveals that the values of the dynamic yield stress of PMMA suspensions are not so negligible, especially for the most concentrated suspension (40 vol % of PMMA) –see Table 1. The non-Newtonian behaviour of the PMMA suspensions is likely related to the existence of some particle interaction different from the hydrodynamic one, which leads to a tendency to aggregation at rest. The existence of particle interactions others than the hydrodynamic one may be checked by comparison of the experimental viscosity of the samples with the theoretical predictions of Krieger-Dougherty equation:[18-19]

$$\eta = \eta_s\left(1 - \frac{\phi}{\phi_m}\right)^{-[\eta]\phi_m} \tag{1}$$

where $\eta_s$ is the viscosity of the liquid carrier, $\phi_m$ is the maximum-packing volume fraction (supposed to be 0.62 for spheres of uniform diameter) and $[\eta]$ is a parameter whose value can be considered as 2.5 for rigid spheres.[18-19] The viscosity calculated by Krieger-Dougherty equation together with the experimental plastic viscosity (obtained as the slope of a linear fit of the CR rheograms for $\dot{\gamma} > 200$ s$^{-1}$) are presented in Table 1 for the PMMA suspensions.

From data in this table it is interesting to note that the deviation from Krieger-Dougherty equation is quite low for the PMMA suspensions with 20 and 30 vol % solid concentration, while it becomes important for the case of 40 vol %. These results suggest that the non-hydrodynamic interaction must be short-ranged since at high volume fractions, the interparticle distance becomes shorter. Given the non-polar nature of the suspending liquid (mineral oil), we expect this interaction to be of van der Waals type.

The most noticeable result evidenced by the data shown in Table 1, however, is the increase of both the viscosity and the yield stress when a fraction of PMMA particles is substituted by iron ones. This effect is not due to an increase of the total volume fraction, since it is the same for every two pairs of suspensions. In fact, it is quite surprising since, according to Barnes et al.,[20] when particles of two different sizes are mixed, the viscosity of the resulting mixture is usually lower than for suspensions with the same volume fraction of monosized particles, as a consequence of the fact that small particles pack into the interstices between the large ones.[20] However, in our particular case, the effect is exactly the opposite: the viscosity increases in the case of the multi-component suspensions. In addition, it is observed that the transition to the flow regime (characterized by non-negligible shear rate values) for the multi-component suspensions takes place in a much more abrupt way (intermediate plateau, see Figure 1b), as it is the case of flocculated suspensions. Therefore, these results seem to indicate that there must be rather strong interactions between the dispersed particles in the multi-component suspensions.

Similar results (increase of yield stress and viscosity) were reported for suspensions constituted by a mixture of organoclay and iron particles in kerosene.[13] The formation of a clay network (gel) that entrapped the iron particles was the reason given in ref.[13] for the high values of the yield stress in the case of organoclay/iron suspensions. However, the possibility of gelformation can be rejected in the case of PMMA particles, as evidenced by the fact that the static yield stress is almost negligible for single-component suspensions of PMMA particles. In order to further investigate particle interactions in the absence of applied field and their effects, the rheological behaviour at zero-field of a suspension of 10 vol % of iron particles (sample S0(10:0)) is also studied. The values of viscosity and yield stress calculated from the rheograms (not shown here for brevity) are presented in Table 1 too. First of all, it is clear from these data that both the dynamic and static yield stresses of sample S0(10:0) are not negligible. This result suggests that the suspension may be quite flocculated. This is not surprising, given that magnetic interaction between iron microparticles is relatively important, even in the absence of external magnetic fields, due to their remnant magnetization (note that due to their micrometric size, these particles are multi-domain from the magnetic viewpoint). In addition to magnetic attraction, van der Waals interaction between the dispersed particles must be fairly important too.

Note that although the yield stress for sample S0(10:0) is quite high despite its low volume fraction, the deviation of its viscosity from the prediction of Krieger-Dougherty equation is not so high. This can be explained because the experimental viscosity is calculated by a linear fit of the high shear rate points of the rheogram ($\dot{\gamma} > 200$ s$^{-1}$). For these values of the shear rate, the forces of attraction between iron particles due to their remnant magnetization and van der Waals interaction are negligible as compared with the hydrodynamic interactions.

Let us note at this point that the rheological parameters (i.e. yield stress, viscosity) of sample S0(10:0) are always lower than those of the multi-component magnetic suspensions. However, there is something that has not been taken into account, namely the relative available space for the movement of the magnetic particles, which is higher in sample S0(10:0) than in the case of the multi-component suspensions. For this reason, and to discard that it is not the lack of available physical space what causes the increase of the yield stress and viscosity in multi-component suspensions, we prepared a suspension of 14 vol % of iron microparticles. In this suspension, the available space for the magnetic particles is the same as in sample S3(10:30). Despite this, and as observed in Table 1, the rheological parameters of the 14 vol % sample are closer to sample S0(10:0) than to sample S3(10:30). Therefore, the high value of the yield stress and the viscosity at zero-field in the multi-component magnetic suspensions is not related exclusively to the non-magnetic particles or to the magnetic ones, but to the combination of both of them. Therefore, there must be some interaction between the two populations of particles, and the purpose of the next section is to find out its nature.

**3.2. Microscopy in the absence of magnetic field and attenuated total reflectance (ATR) observations**

With the aim of investigating if the two populations of particles in the multi-component suspensions indeed interact, microscopic observations of the dispersed particles in dilutions of the original suspensions were obtained by using Haake MARS III Rheoscope Module (Thermo Fisher Scientific, USA). It was necessary to dilute the concentrated samples (by pouring aliquots of them into appropriate amounts of mineral oil) in order to let the light pass through the sample. Figure 2 shows a photograph of a 1:5 dilution of sample S1(10:10). Similar results were obtained for other dilution ratios and/or original samples.

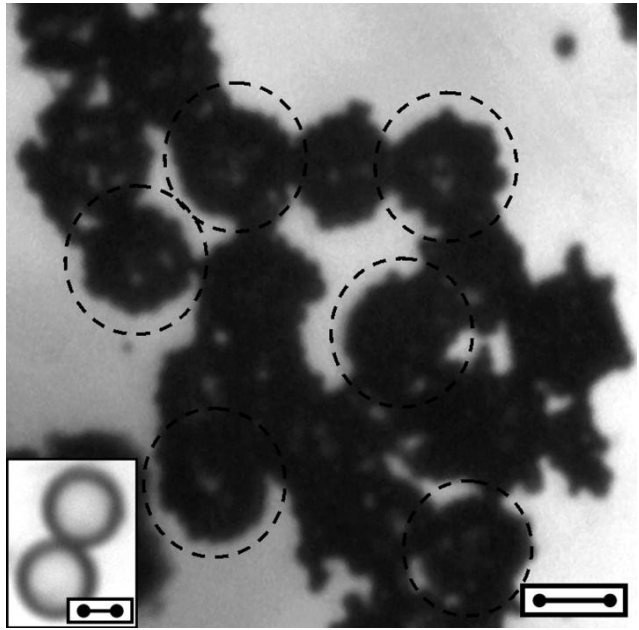

**Fig. 2** Microscopic picture of a dilute iron/PMMA sample (bar length: 10 microns). Dilution ratio 1:5 as compared to the original suspension S1(10:10). Note that there are several PMMA particles almost completely coated by iron ones (some of them have been highlighted by dash circumferences). Inset: microscopic picture of two bare PMMA particles for comparison (bar length: 5 microns).

It is seen that in the absence of magnetic field, PMMA particles (some of which are highlighted by a dash line in Figure 2) are almost completely coated by a shell of iron ones. This makes evident that there are interactions between both populations of particles. These interactions lead to some kind of non-magnetic-core–magnetic-shell composites (as seen in Figure 2), which are likely responsible for the increase of both the viscosity and the yield stress in the absence of field. The affinity between PMMA and iron is not surprising, since PMMA has been previously used to stabilize colloidal iron particles.[21-25] For example, Guo et al.[22] described the preparation of ferromagnetic composites consisting of iron core and PMMA shell.

With the purpose of gaining further information on the interaction between the magnetic and the non-magnetic particles, we performed attenuated total reflectance (ATR) observations with an infrared spectrometer JASCO 6200 (Japan). The spectrum of sample S3(10:30) and the one of a sample with the same volume fraction (30 %) of PMMA are shown in Figure 3 as an example.

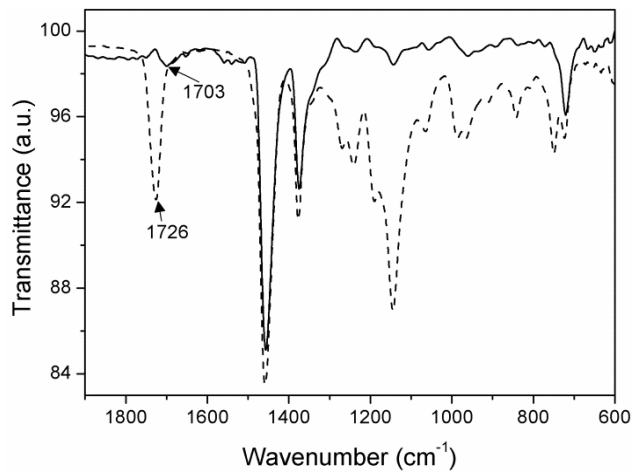

**Fig. 3** ATR spectra of a sample consisting of 30 vol % of PMMA dispersed in mineral oil (dash line) and of sample S3(10:30) (solid line).

The peaks at around 1452, 1375 and 720 cm$^{-1}$ in Figure 3 correspond to mineral oil. The peaks at 1145 and 1191 cm$^{-1}$ are related to C-H deformations, and the ones at 1241 and 1271 cm$^{-1}$ to C-C-O stretch coupled with C-O stretch.[22] Other characteristic absorption vibration peaks of PMMA appear at 838, 984 and 1052 cm$^{-1}$. The characteristic stretching band of the carbonyl group (C=O) of PMMA appears at 1726 cm$^{-1}$ for the 30 vol % PMMA suspension. However, in the case of sample S3(10:30) it is displaced to a lower wavenumber (1703 cm$^{-1}$). Such a change is due to interaction between the surface of PMMA and the iron particles. More specifically, it has been reported that when the carbonyl group of carboxylic acids and derivatives is complexed by dative coordination, its characteristic stretching band is shifted to lower frequencies (i.e. lower wavenumbers).[22,24,26-29] For example, a shift of the carbonyl stretching band from 1732 cm$^{-1}$ (pure PMMA) to 1706 cm$^{-1}$ (PMMA-stabilized iron nanoparticles) was taken as a proof of the existence of chemical bonding between PMMA and iron nanoparticles in a recent work by Guo et al.[22] The formation of coordination compounds between organic ligands and the surface sites of iron (or iron oxides) has also been widely studied.[30]

**3.3. Steady state rheological measurements upon magnetic field application**

In this section, the rheological behaviour of the samples in the presence of external magnetic fields is studied. These measurements were performed similarly to those of section 3.1, but this time, a homogeneous magnetic field was applied in the vertical direction (that is, perpendicular to the rheometer plates) by means of a coil. The field was applied since the beginning of the waiting time. Figure 4 shows the rheograms obtained in the CR (a,c) and CS (b,d) modes at different values of the internal magnetic field, *H,* for samples S0(10:0) and S2(10:20) as an example. Note that the internal magnetic field was calculated as $H_0/\mu_r$, where $H_0$ is the applied magnetic field and $\mu_r$ the relative magnetic permeability of the sample. $H_0$ was experimentally determined with the help of a gaussmeter, whereas the values of $\mu_r$ for each applied magnetic field were obtained by a Squid Quantum Design MPMS XL magnetometer.

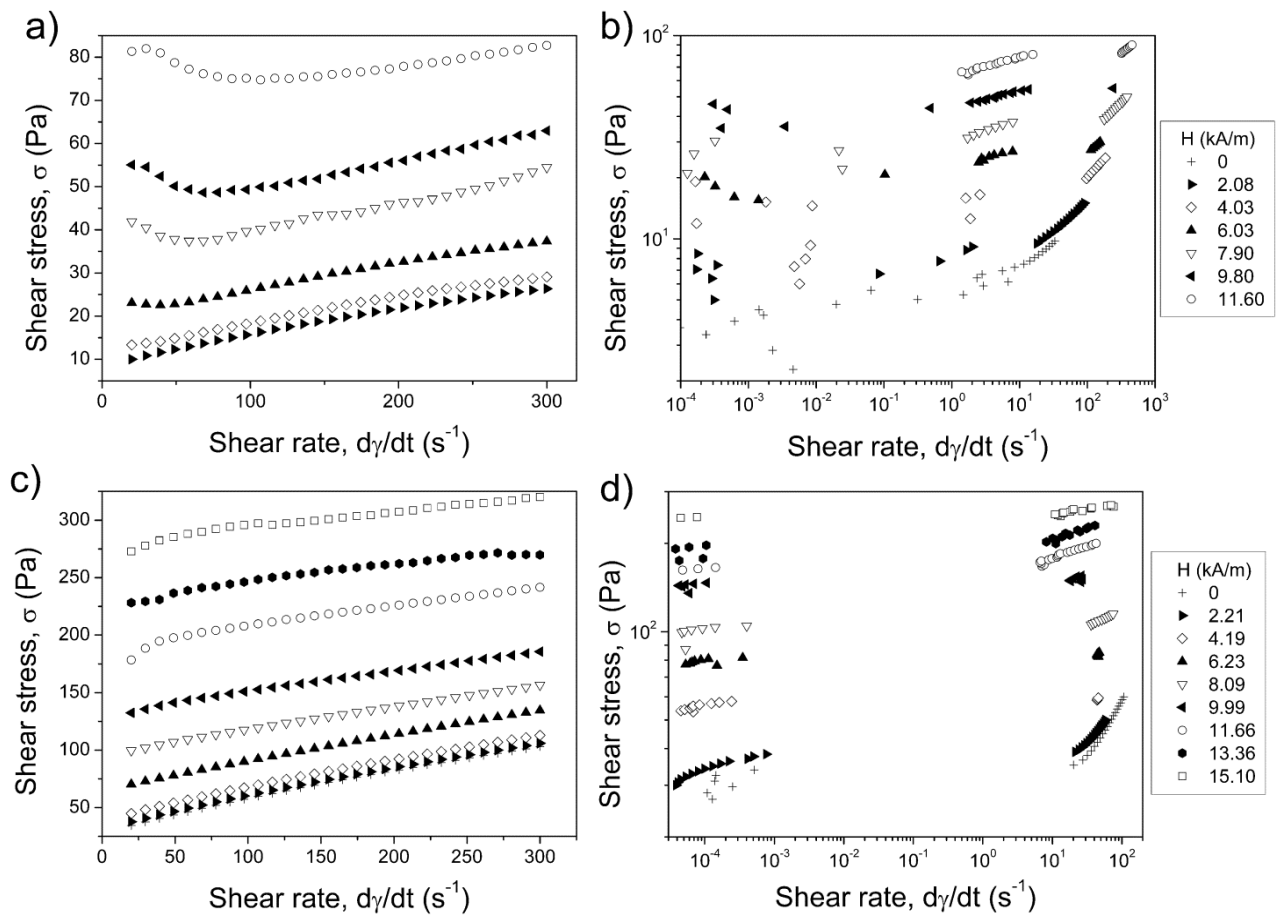

**Fig. 4** CR (a,c) and CS (b,d) rheograms for samples S0(10:0) (a,b) and S2(10:20) (c,d). The different curves correspond to different values of the internal magnetic field.

With reference to the results in Figure 4 it is seen that there is a clear MR effect for both samples, that is, an increase of the shear stress when the magnetic field strength is increased for a particular shear rate. Therefore our multi-component suspensions respond to the magnetic field in a similar way than single-component (iron) MR suspensions. However, the MR effect is much stronger for sample S2(10:20) than for sample S0(10:0), as evidenced by the fact that the shear stress at similar internal magnetic fields is higher for the former, despite containing both suspensions the same amount of magnetic particles. This is in good agreement with the results presented in refs. 14-16. From Figure 4 it is also interesting to highlight the yielding behaviour of these suspensions, characterized by the appearance of a yield stress. Such behaviour is especially noticeable in the CS rheograms (Figures 4b and 4d), for which an abrupt transition from negligible shear rate values to the flow regime appears. Note that this abrupt transition is not observed in the CR measurements, since in this case, the shear rate is stepwise imposed by the rheometer. In order to gain more information on the MR response of the samples, both the dynamic and static yield stresses were obtained from curves like those shown in Figure 4. In Figure 5 the increments of both quantities with respect to their values in the absence of applied field are plotted against the internal magnetic field for all the samples.

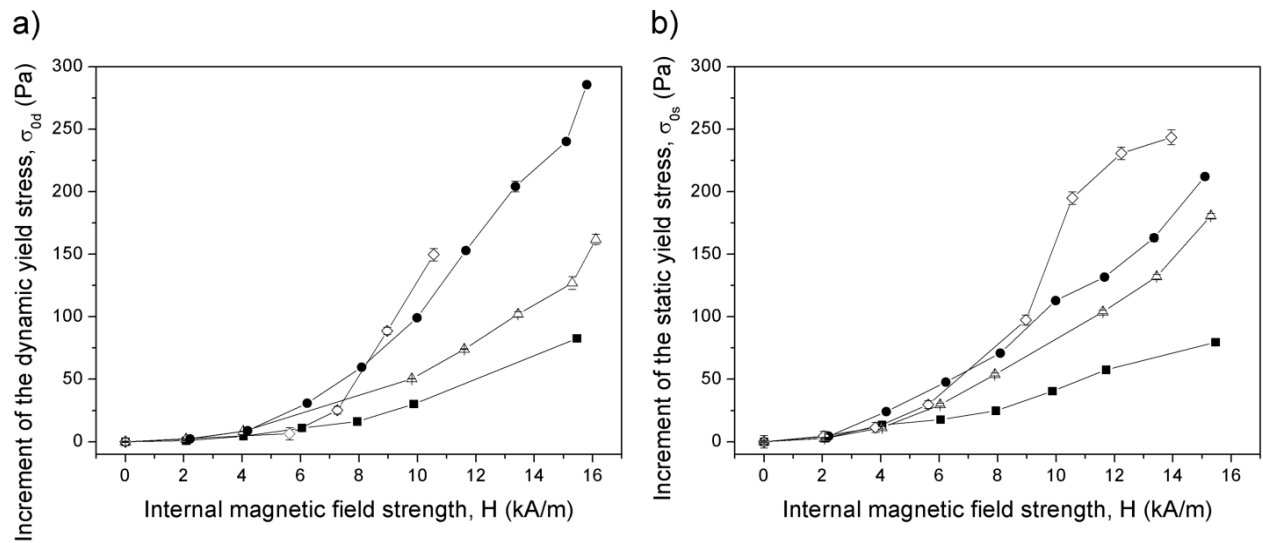

**Fig. 5** Magnetic field dependence of the dynamic (a) and static (b) yield stress increments for all the samples. ■: S0(10:0); △: S1(10:10); ●: S2(10:20); ◇: S3(10:30). Solid lines represent eye guidelines.

As observed in Figure 5, both the static and dynamic yield stress increments increase with the magnetic field for all the samples, as is usually the case of MR fluids. Moreover, the static yield stress is, in general, lower than the dynamic yield stress. Since the static and dynamic yield stresses are related, respectively, to the stresses needed to start the flow and to break all the structures in the suspension, it is logical that the first one is slightly lower than the latter.

Nevertheless, what seems more interesting is that the sample with the lowest values of the yield stress is the one without PMMA particles. In fact, both the dynamic and the static yield stresses increase with the concentration of non-magnetic particles, with the exception of sample S3(10:30), which exhibits a quite irregular behaviour, since, at low internal fields its yield stress is lower than the one of sample S2(10:20), while this behaviour reverses at higher internal magnetic fields.

### 3.4. Microscopy under magnetic field

Microscopic observations upon magnetic field of dilutions of the multi-component suspensions were performed in order to examine the field-induced particle chaining. The observations were carried out by placing Haake MARS III Rheoscope Module between two Helmholtz coils that were used for the application of a magnetic field parallel to the surfaces that confined the sample. Note that the magnetic field was homogeneous near the middle of the common axis of the coils, where the samples were placed for observation (variation of the magnetic induction of less than 1%). The magnetic field was raised stepwise until the maximum field applied by the coils, 9.8 kA/m, was reached. The original samples were diluted as described in paragraph 3.2.

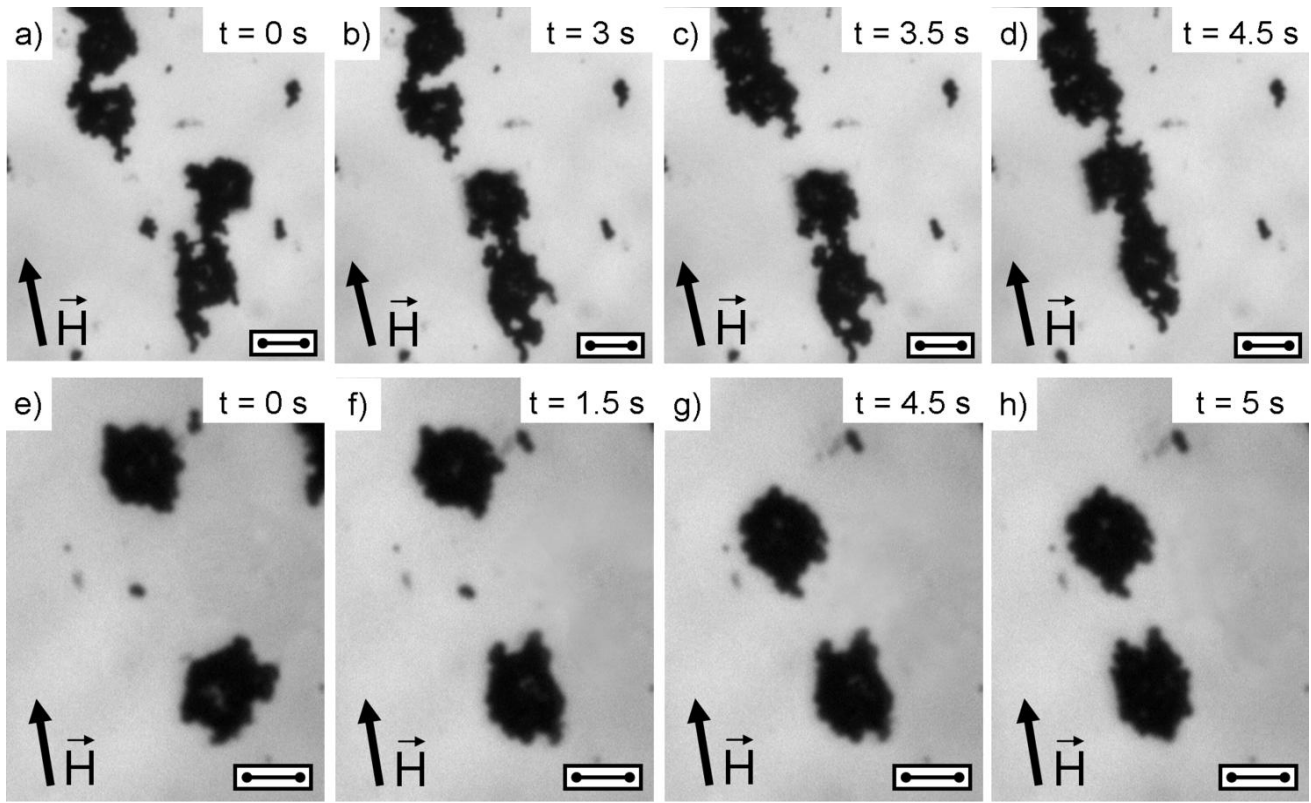

**Fig. 6** Sets of photographs for dilute iron/PMMA samples. Dilution ratio as compared to the original suspension S1(10:10) is 1:25. (a-d) and (e-h), represent observations of different aliquots. Figures (a) and (e) were taken prior to the application of the magnetic field while the rest of them correspond to different moments (indicated in seconds in the photographs) after turning on a magnetic field of 9.8 kA/m of strength. The arrows indicate the direction of the applied magnetic field.

The sets of photographs shown in Figure 6 are characteristic of the observed particle behaviour in the dilute multi-component suspensions upon field application –see also Video 1. Let us first focus on the photographs of Figures 6a-6d, which correspond to a time sequence of the same region. In the absence of magnetic field (Figure 6a) PMMA particles appears coated by a thick shell of iron particles, in agreement with results of section 3.2. As seen in Figures 6b-6d, when the magnetic field is switched on, the iron/PMMA composites begin to move and show a tendency to form aggregates orientated in the field direction. This is due to the adsorbed shell of iron particles, which reacts to the application of the magnetic field. The result is the formation of field-induced particle structures constituted by non-magnetic-core–magnetic-shell composites. Figures 6e-6h show the approximation of two of those composites upon field application.

In view of these results, it seems that the enhancement of the MR response of the multi-component suspensions could be related to the non-magnetic-core–magnetic-shell nature of the iron/PMMA composites. From classic magnetism it is known that hollow magnetisable spheres have a higher magnetic moment than solid, continuous ones with the same amount of magnetic material. Based on the analytical solution of the magnetostatic problem,[31-32] it is easy to show that, in the limit of high magnetic permeability, $\mu_r \gg 1$, the magnetic moment of an isolated hollow sphere, $m_h$, is approximately equal to that of a full sphere, $m_f$, divided by the volume fraction $\phi_{mag}$, of magnetic material in the hollow sphere: $m_h = m_f/\phi_{mag}$. This indicates that thinner shells contribute to a stronger magnetic response of the suspension. Therefore, we expect a similar relationship for the magnetic susceptibility of a suspension of non-magnetic-core– magnetic-shell particles. In our case, the thickness of the iron shell around the PMMA particles should progressively decrease with increasing concentrations of PMMA particles at a constant volume fraction of iron ones. This qualitatively explains the increase of the magnetic (and therefore MR) response of the suspension with the volume fraction of non-magnetic particles. However, this analysis is only valid for non-structured, isotropic and very dilute suspensions. Numerical simulations by finite element methods (FEM) are required for a more precise comparison of the strongly structured concentrated suspensions. This is the purpose of the next section.

**4. Theory**

In this section, the static yield stress of the suspensions and its increase when PMMA and iron particles are dispersed together, are estimated by the use of a theoretical model first developed to study the static yield stress of concentrated suspensions of multi-domain magnetic particles confined between two parallel surfaces.[33] In this model, the magnetically-induced particle clusters are considered to be body-centred tetragonal (BCT) structures, with a central particle chain surrounded by four peripheral ones shifted vertically by a particle radius (see Figure 7). These structures have proved to be the most favourable from the energetic viewpoint.[34] Besides, they have been previously observed by laser diffraction in electrorheological fluids –the electric counterparts of magnetorheological suspensions.[35] At zero strain, the neighboring particles of a chain are in contact, but when sheared, the structure is stretched along its major axis and the particles of the central chain are separated from each other (at a distance given by an affine deformation). In addition, the particles belonging to the peripheral chains are pulled inside the cluster so that they keep contacts with the central chain, as depicted in Figure 7. Note that this structure is expected to be mechanically stable at the experimental time scale, especially when compared to the single chain structures used in some traditional models (for more details, see discussion in ref.[33]).

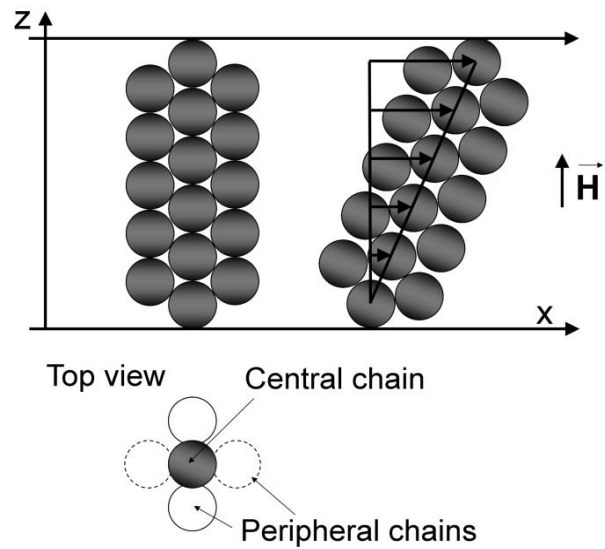

**Fig. 7** BCT structures considered in the model for the static yield stress presented in ref.[33]. Left: structure at rest; right: structure upon shear; down: top view of the structure.

The structures are supposed to be always stuck to both the upper and the lower plates of the rheometer, since highly rough surfaces were employed in order to avoid wall slip. When the upper plate is displaced a distance $\Delta x$, a strain of magnitude $\gamma = \Delta x / h$ is applied to the structure ($h$ is the gap between the plates). The structure is homogeneously deformed until its failure. The deformation of the BCT structures of Figure 7 occurs in two ways: i) they are tilted when they are pulled by the upper plate. As a consequence, a restoring magnetic torque that tends to align them back to the external magnetic field appears; (ii) the BCT-chains are stretched and extended along their major axis, and their constituent particles are separated from each others, giving rise to interparticle gaps and restoring forces along the main axis of the structure (see Figure 7). Under homogeneous deformation, the main axes of magnetization of the suspension coincide with the axes of the BCT structure and therefore, when a reference frame linked to the structure is considered, the magnetic permeability tensor is diagonal, and only has two dissimilar components: $\mu_{//}$ and $\mu_{\perp}$. These two components correspond to the major and minor axes of the structure and are referred to as longitudinal and transversal magnetic permeability respectively.

In the limit of particles with high magnetic permeability, the theoretical static yield stress can be calculated by the following equation according to ref.[33]:

$$\sigma \cong -\frac{1}{2}\mu_0 H^2 \left[ \frac{\partial \mu_{//}}{\partial \gamma} \cdot \frac{1}{1+\gamma^2} + \frac{\partial \mu_{\perp}}{\partial \gamma} \cdot \frac{\gamma^2}{1+\gamma^2} \right] \quad (2)$$

where $\mu_0$ is the magnetic permeability of vacuum. The permeability components, $\mu_{//}$ and $\mu_{\perp}$ are calculated from the magnetic field distribution in the suspension, using the following expressions:

$$\mu_{//,\perp} = \frac{1}{\mu_0 H \cdot V} \int B_{//,\perp} dV \quad (3)$$

where $\int B_{//,\perp} dV$ is the integral of the magnetic flux density over the suspension volume ($V$). This integral is calculated by solving the Maxwell equations for the magnetostatic potential, $\Psi$ (introduced by $\mathbf{B}=\mathrm{rot}\Psi$) through finite element method simulations with the help of the software FEMM.[36] For this purpose we consider a parallelepiped representative cell containing a multi-chain BCT structure. The central chain consists of six spherical particles (Figures 8a and 8c). Different strains, $\gamma$, are imposed by separating the central chain particles at a relative distance equal to $\delta/a = 2(\sqrt{1+\gamma^2}-1)$ where $a$ is the particle radius. At the same time the particles of the peripheral chains are approached in such a way that there is no fracture of the structures. Since for stress calculations we only need the diagonal components of the magnetic permeability tensor, $\mu_{//}$ and $\mu_{\perp}$, we can consider only two directions of the magnetic field: longitudinal (Figure 8a) and transverse (Figure 8c) with respect to the main axis of the BCT cluster. The applied boundary conditions are specified in the caption of Figure 8.

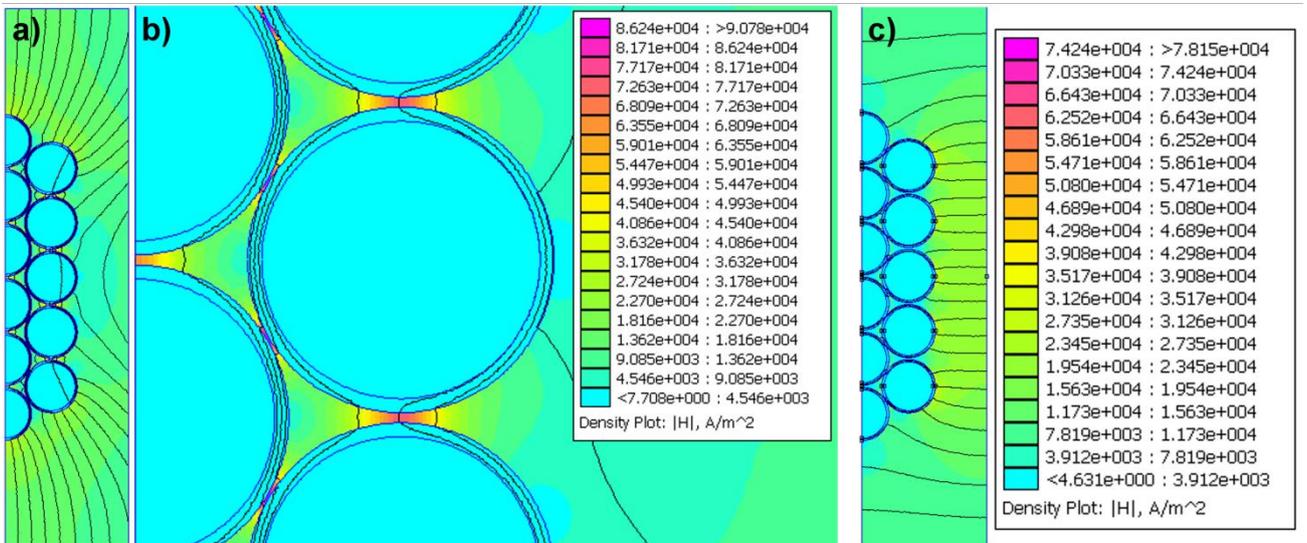

**Fig. 8** Magnetic field intensity distribution in the unit cells for FEMM simulation of sample S3(10:30). Two directions of the magnetic field are considered for the estimation of the magnetic permeability tensor components: longitudinal (a) and transverse (c). (b) is a magnification of (a) where the enhancement of the field in the interparticle gaps is clearly observed. Boundary conditions are as follows: (a) left boundary $\Psi = 0$ (plane of symmetry); right boundary $(1/\mu_0)\partial\Psi/\partial n = H = 10\ kA/m$ (mean field in the suspension); upper and lower boundaries $\partial\Psi/\partial n = 0$ (the magnetic field is longitudinal far from the clusters); (c) left and right boundaries $\partial\Psi/\partial n = 0$ (magnetic field is perpendicular due to symmetry reasons); upper and lower boundaries $(1/\mu_0)\partial\Psi/\partial n = H = \pm 10\ kA/m$.

The simulations were performed for two different magnetically-induced structures: (i) of just iron particles (sample S0(10:0)) and (ii) of PMMA particles coated by a uniform layer of iron (samples S1(10:10), S2(10:20) and S3(10:30)). The dimensions of the cells were adjusted so that the volume fractions of the different materials in the cells are the same as in the real suspensions. Note that the consideration of a uniform layer of iron around a non-magnetic particle is not completely realistic, since it has been previously observed that the magnetic coating is not totally uniform. Such simplification may give rise to quantitative differences between the experimental data and the theoretical results. However, as we will see, it gives a quite reasonable approach that allows predicting the observed trends, which justifies this approximation.

Figure 8 shows the magnetic field intensity distribution for the BCT structures in sample S3(10:30), as an example, for the two directions of the magnetic field: longitudinal (Figure 8a) and transverse (Figure 8c). Figure 8b is a magnification of Figure 8a. As observed, the magnetic field intensity is strongly enhanced in the gaps between the core-shell PMMA-iron composites. This result reveals the importance of the interparticle gaps as the main source of the stress in concentrated suspensions of magnetic particles upon magnetic field.

Figure 9 shows the strain dependence of the longitudinal and transverse magnetic permeabilities, obtained by eq. 3, for samples S0(10:0) and S3(10:30) at an internal magnetic field of 10 kA/m, as an example. It is seen that $\mu_\perp$ is almost independent of the strain. This is because $\mu_\perp$ depends mainly on the distance between particles of two opposite pheripheral chains in the BCT structures. When the structures are sheared, the particles of these chains approach, but the distance between them scarcely changes with respect to its initial value. The tendency for $\mu_\parallel$ is completely different, with an important decrease for increasing values of the strain. This is due to the appearance and enlargement of gaps between particles of the same chain, when the structures are deformed. However, the most noticeable result of Figure 9 is the increase of both components, $\mu_\parallel$ and $\mu_\perp$ when PMMA particles are added to the suspension.

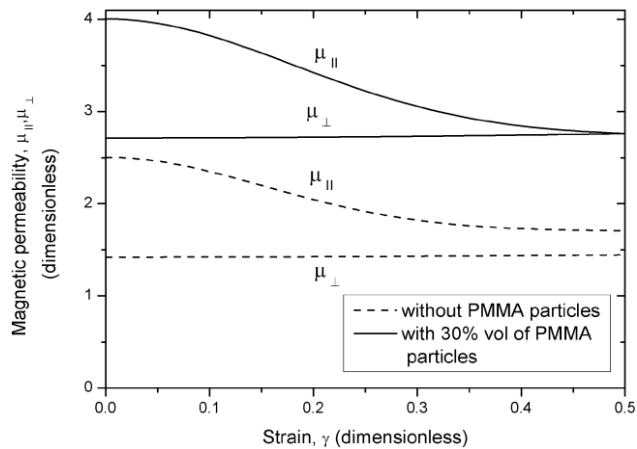

**Fig. 9** Dependence on the strain of the longitudinal ($\mu_\parallel$) and transverse ($\mu_\perp$) magnetic permeabilities at an internal magnetic field of 10 kA/m for samples S0(10:0) (dash line) and S3(10:30) (solid line).

The shear stress calculated by using eq. 2 at an internal magnetic field of 10 kA/m is shown in Figure 10 for samples S0(10:0) and S3(10:30) as an example. It is seen that the shear stress remains higher in the case of sample S3(10:30) for all the range of strain values.

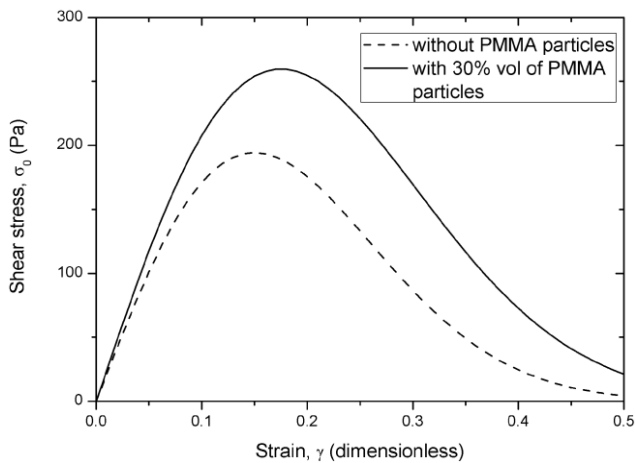

**Fig. 10** Calculated shear stress as a function of the strain at an internal magnetic field of 10 kA/m for samples S0(10:0) (dash line) and S3(10:30) (solid line).

As mentioned in section 3.1, the static yield stress can be experimentally estimated by extrapolating the value of the stress in the low shear rate pseudoplateau of double logarithmic scale rheograms. Alternatively, it can be calculated as the maximum value of the shear stress in shear stress vs strain curves like those shown in Figure 10. Above the critical value of the shear strain (i.e. the one for which the maximum appears), particle structures break and give rise to a decrease of the shear stress. In Figure 11 the so-obtained static yield stress is plotted against the internal magnetic field together with the experimental results for samples S0(10:0) and S3(10:30). As seen in this figure, the theoretical model correctly predicts the increase of the static yield stress of sample S3(10:30) with respect to sample S1(10:0). However, it overestimates the static yield stress in both cases. This deviation from the experimental data can be attributed to the simplifications made in our model (e.g. homogeneous iron shell), and to the fact that the real particle structures in the suspensions may be slightly different from those considered in our model (see ref.[37] for discussion on field-induced particle structures in MR fluids). However, the model correctly predicts the experimentally-observed trends. It is important to remark that it is the increase of the magnetic permeability of suspensions of non-magnetic-core–magnetic-shell composites with respect to suspensions of solid (pure) magnetic particles, which qualitatively explains the increase of the yield stress of the multi-component magnetic suspensions.

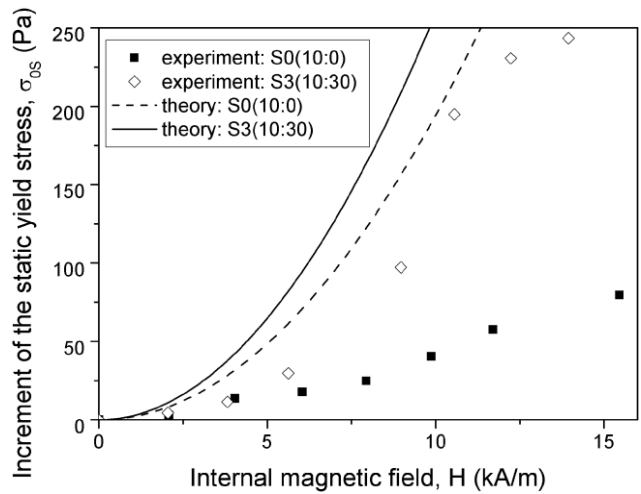

**Fig. 11** Theoretical values (lines) and experimental increment (symbols) of the static yield stress. Dash line and ■ symbols correspond to sample S0(10:0) and solid line and ◇ symbols to sample S3(10:30).

**Conclusions**

In this work, the steady state rheological behaviour in the absence of magnetic field and upon its application of a kind of multi-component magnetic suspensions has been studied. These suspensions consist of two particle populations: a magnetic one, consisting of iron particles with median diameter of 2.3 μm, and another of non-magnetic, PMMA particles of about 10 μm in diameter. The iron:PMMA volume fractions in the suspensions have been set to 10:0, 10:10, 10:20 and 10:30. In addition, the behaviour of single-component suspensions consisting just of PMMA (20, 30 and 40 vol %) and iron (10 and 14 vol %) has been also studied. It has been observed that at zero-field, the multi-component suspensions develop higher values of the viscosity and the yield stress than the corresponding single-component ones in spite of having the same total volume fraction. Such an increase has been proved not to be due to interactions between particles of the same population, but to the interaction between iron and PMMA particles. More specifically, it has been explained by the appearance of core-shell composites, which arise from the adsorption (i.e. coordination compound formation) of iron particles on PMMA ones.

The existence of such an adsorbed iron-on-PMMA layer has been confirmed by optical microscopy, both in the absence of magnetic field and upon its application as well as by ATR. Furthermore, upon magnetic field application, the resulting non-magnetic-core–magnetic-shell composites move and present a tendency to aggregate into structures aligned in the direction of the applied field. It has been shown that these magnetic field-induced structures, which involve both particle populations, give rise to an increase of the suspension viscosity and dynamic and static yield stresses upon magnetic field application, with respect to the single-component suspension with the same volume fraction of iron (sample S0(10:0)). In fact, it has also been observed that both yield stresses increase when the PMMA volume fraction increases. As a result it can be said that the incorporation of non-magnetic particles into a suspension of magnetic ones, gives rise to an enhanced MR effect. Finally, a theoretical model has been applied in order to explain the origin of this enhancement of the MR effect. According to this model, the main contribution to the yield stress is the change of the suspension magnetic permeability when the magnetic field-induced structures are stretched by the shear. By means of finite element method simulations it has been showed that the magnetic permeability of multi-component suspensions is higher than this of single-component suspensions having the same amount of magnetic particles. Consequently, this increase of the magnetic permeability is the ultimate reason for the experimentally observed enhancement of the MR effect when non-magnetic particles are coated by a layer of magnetic material.

**Acknowledgements**

This work has been supported by Projects P08-FQM-3993, P09-FQM-4787 (Junta de Andalucía, Spain) and FIS2009-07321 (MICINN, Spain). In addition, L. Rodríguez-Arco and M.T. López-López acknowledge financial support by Secretaría de Estado de Educación, Formación Profesional y Universidades (MECD, Spain) through its FPU program and University of Granada (Spain) respectively.